\documentclass[
reprint,
amsmath,amssymb,
aps,
]{revtex4-2}
\usepackage{graphicx}

\begin{document}

\title{Great year, bad Sharpe?\\ A note on the joint distribution of performance and risk-adjusted return}

\author{Matteo Smerlak}
\affiliation{Laboratoire Biophysique et Evolution, ESPCI, 10 rue Vauquelin, F-75005 Paris}

\date{\today}

\begin{abstract}
    Returns distributions are heavy-tailed across asset classes. 
    In this note, I examine the implications of this well-known stylized fact for the joint statistics of performance (absolute return) and Sharpe ratio (risk-adjusted return). 
    Using both synthetic and real data, I show that, all other things being equal, the investments with the best in-sample performance are never associated with the best in-sample Sharpe ratios (and vice versa). 
    This counter-intuitive effect is unrelated to the risk-return tradeoff familiar from portfolio theory: it is, rather, a consequence of asymptotic correlations between the sample mean and sample standard deviation of heavy-tailed variables. 
    In addition to its large sample noise, this non-monotonic association of the Sharpe ratio with performance puts into question its status as the gold standard metric of investment quality. 
 \end{abstract}

\maketitle 

\section{Introduction}

Of two investments with comparable returns, the one with lower volatility is more desirable. 
From this commonsense observation derives the notion of risk-adjusted returns as target for optimization. 
This logic is formalized in Markowitz's ``modern portfolio theory'', wherein a portfolio is considered efficient if it maximizes the Sharpe ratio (mean return)/(volatility). 
In practice, Sharpe ratios or similar risk-adjusted performance measures are routinely used to benchmark and select funds, compensate their managers, etc.

From a statistical perspective, the Sharpe ratio is simply the coefficient of variation of excess returns. 
Assuming returns with finite variance, its sampling distribution can be described with standard asymptotic theory \cite{loStatistics2002}. 
However, precisely because investors are focused on maximizing their Sharpe ratio, it interesting to ask not just about the center of its sampling distribution, but also about its tails. 
What kind of events lead to large deviations of the Sharpe ratio? 

The purpose of this note is to examine the tails of the joint sampling distribution of performance and Sharpe ratio of funds.
A natural expectation is that the Sharpe ratio takes the largest values along trajectories with exceptionally large returns.
I show that this expectation would be true in a Gaussian world, but is false when returns are heavy-tailed \cite{bouchaudTheory2003}. 
In particular, I show that, with heavy-tailed returns, the strategies with the largest Sharpe ratios never maximize long-term performance, and strategies with exceptional performance are always associated with suboptimal Sharpe ratios. 

Two cautionary remarks. 
First, the present study is strictly about in-sample statistics: we do not consider issues of prediction or out-of-sample validity.  
Second, the relationship between performance and risk-adjusted returns discussed here is unrelated to the risk-return tradeoff familiar from portfolio theory: throught the paper, the distribution of returns (and in particular its mean $\mu$ and standard deviation $\sigma$) is fixed. 

\section{Results}

\subsection{Definitions}

Let us begin with some definitions and assumptions.
Consider an asset or strategy with price $p_t$ and log-returns at time resolution $\tau$ by $r_t = \log(p_t/p_{t-\tau})$.
Suppose that a riskfree asset provides returns at rate $r_0$, and denote $\eta_t = r_t - \xi_0$ the excess log-return. 
(Henceforth $\eta_t$ will be called \emph{return} for short.)
Assume furthermore that $\eta_t$ are independent and drawn from a common distribution $\mathcal{D}$ with mean $\mu$ and standard deviation $\sigma$. 

\begin{figure}[t!]
    \includegraphics[width = .45\textwidth]{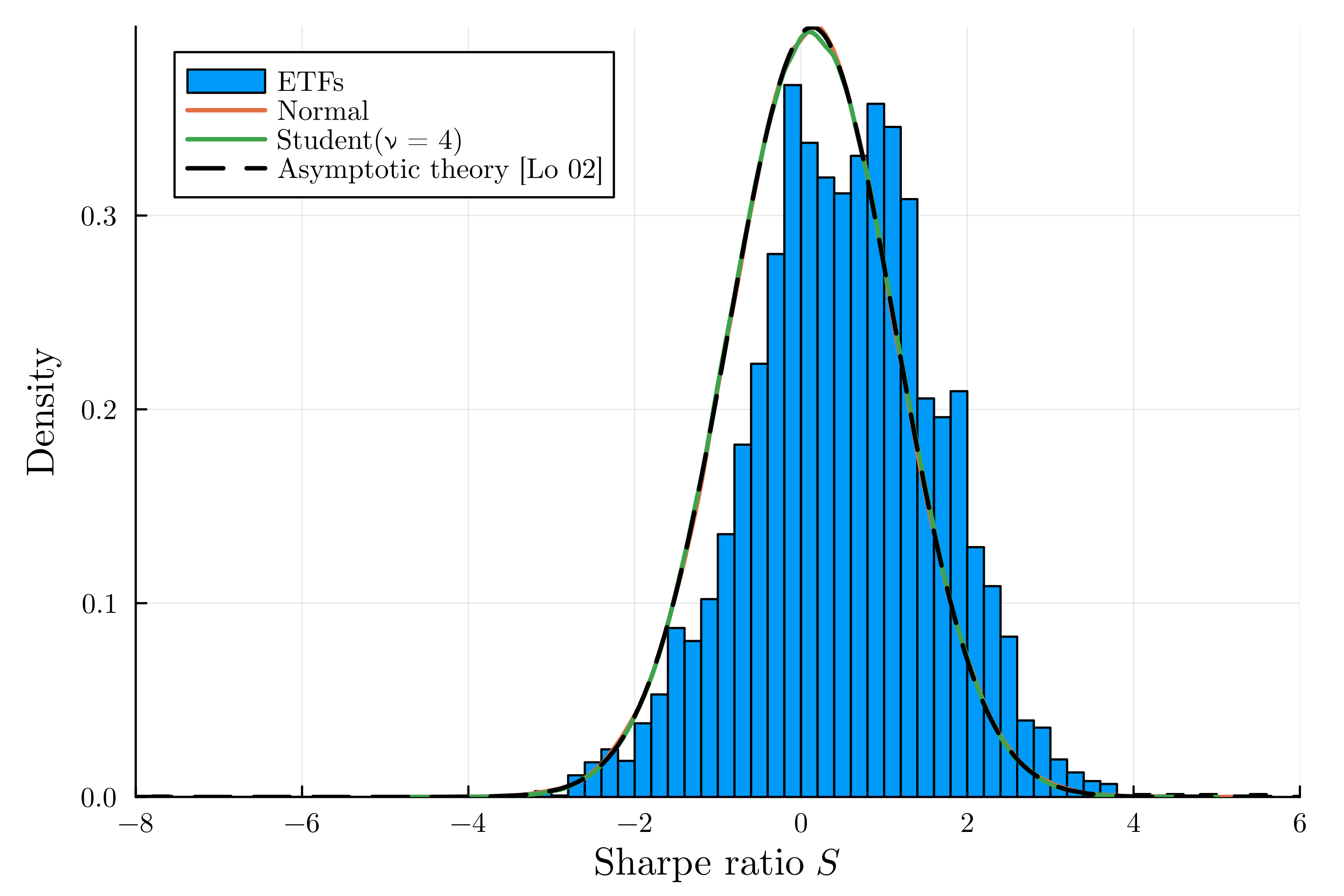}
    \caption{
        Distribution of 1-year Sharpe ratios within the ETF dataset and with synthetic data (Student or Gaussian, both with theoretical $\overline{S} = 0.13$), compared to asymptotic theory \cite{loStatistics2002}. 
        Note the very large standard error $\Delta S = 1.00 \gg \overline{S}$.
        }
    \label{sharpe-dist}
\end{figure}

The performance of the strategy over the time horizon of $T$ periods is measured by its \emph{mean return} $m = \sum_{t=1}^T \eta_{t}/T$: after a time $T$, $1$ dollar invested in this strategy has grown into $R = \exp(mT)$ times the riskfree return $e^{r_0T}$. 
Similarly, the \emph{volatility} of the strategy over the same horizon is the standard deviation of returns, defined by $s^2 = \sum_{t=1}^T (\eta_{t} - m)^2/T$.
Given its performance $m$ and volatility $s$, the \emph{Sharpe ratio} of the strategy is then defined as 
\begin{equation}
    S = \sqrt{T}\, \frac{m}{s}.    
\end{equation}
This ratio is approximately invariant with respect to $T$, and can be interpreted as a signal-to-noise ratio (signal $=mT$, noise $=s\sqrt{T}$). When $T$ is large ($T = 252$ for daily returns), the sampling distribution of $S$ is asymptotically Gaussian, with standard error $\Delta S = (1+S^2/2)^{1/2}$ \cite{loStatistics2002} (see Fig. \ref{sharpe-dist}).

Related measures of risk-adjusted return include the information ratio (where the reference rate $r_0$ is the market rate), the Sortino ratio (where $s$ is replaced by the downside risk), and the Treynor ratio (where $s$ is replaced by $\beta$). These give similar results and we do not discuss them here.

\begin{figure}[t!]
    \includegraphics[width = .45\textwidth]{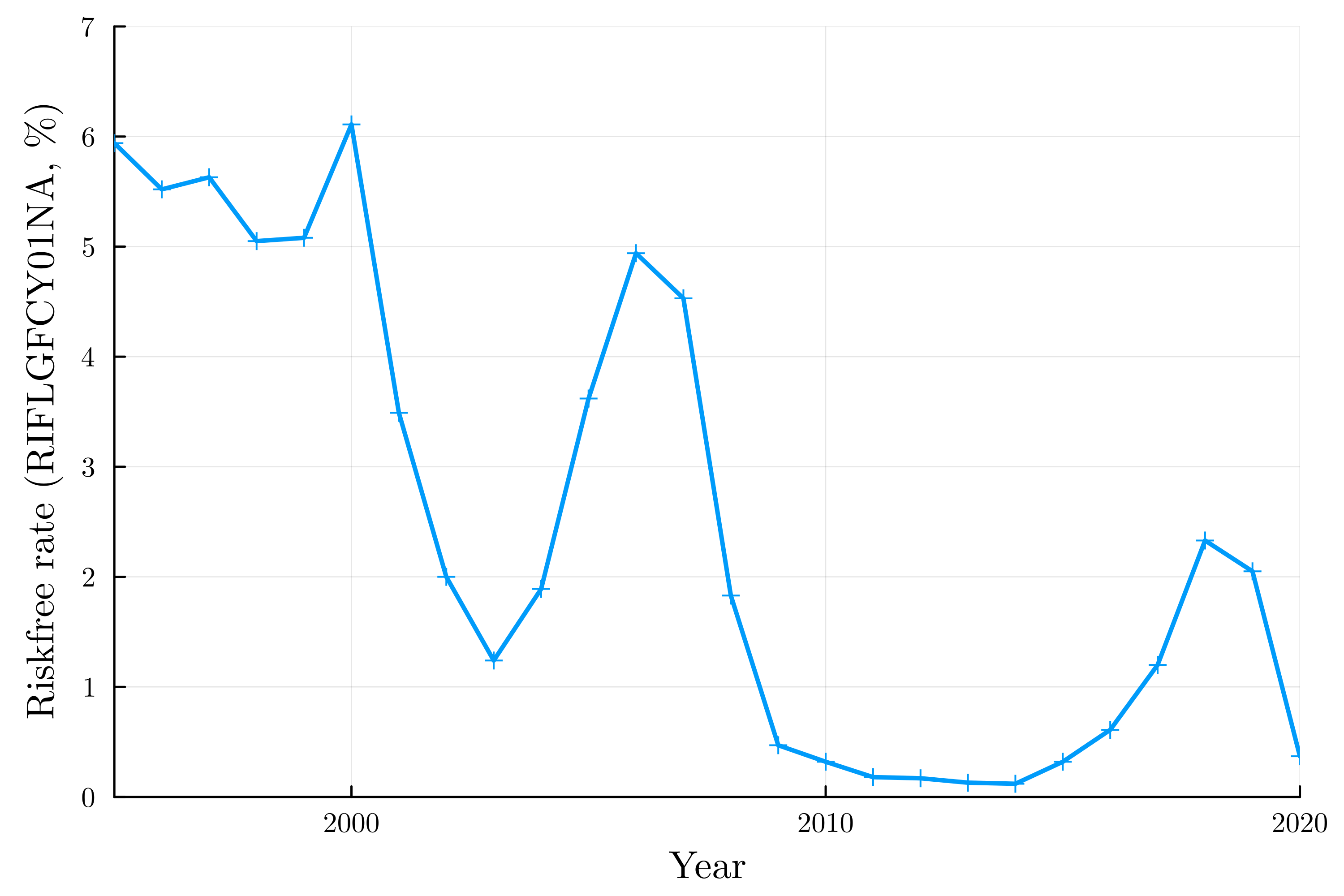}
    \caption{The riskfree rate used to compute the excess returns $\eta_t$ of over 1600 ETFs over the period 1995-2019.}
    \label{riskfree-rate}
\end{figure}

\subsection{Data}

We consider daily returns data of two kinds. First, we generate synthetic returns from a \emph{Student distribution} with mean $\mu$, standard deviation $\sigma$, and tail index $\nu$:
\begin{equation}
    \eta \sim \mu + \sigma\sqrt{(\nu - 2)/\nu}\, \xi,
\end{equation}
where $\xi$ follows a Student $t$ distribution with $\nu$ degrees of freedom. (The associated probability density function reads has tail behavior $p(\eta; \nu)\sim_{\eta\to\pm\infty} 1/\vert\eta\vert^{\nu + 1)}$.) 
Empirically, the tail index of daily returns is found to be $\nu\simeq 4$ (see Ref. \cite{bouchaudTheory2003} and references therein); in the limit $\nu\to\infty$ a Gaussian distribution is recovered. 

We also consider a public dataset of the adjusted close price of exchange traded funds (ETFs) spanning the period 1995-2019 and multiple categories (equities, commodities, short-term bonds, etc.) \cite{leoneUS}. 
To compute excess returns we use as riskfree rate $r_0$ the market yield on 1-year T-bills (the RIFLGFCY01NA ticker from FRED), plotted in Fig. \ref{riskfree-rate}. 

\begin{figure}[t!]
    \includegraphics[width = .45\textwidth]{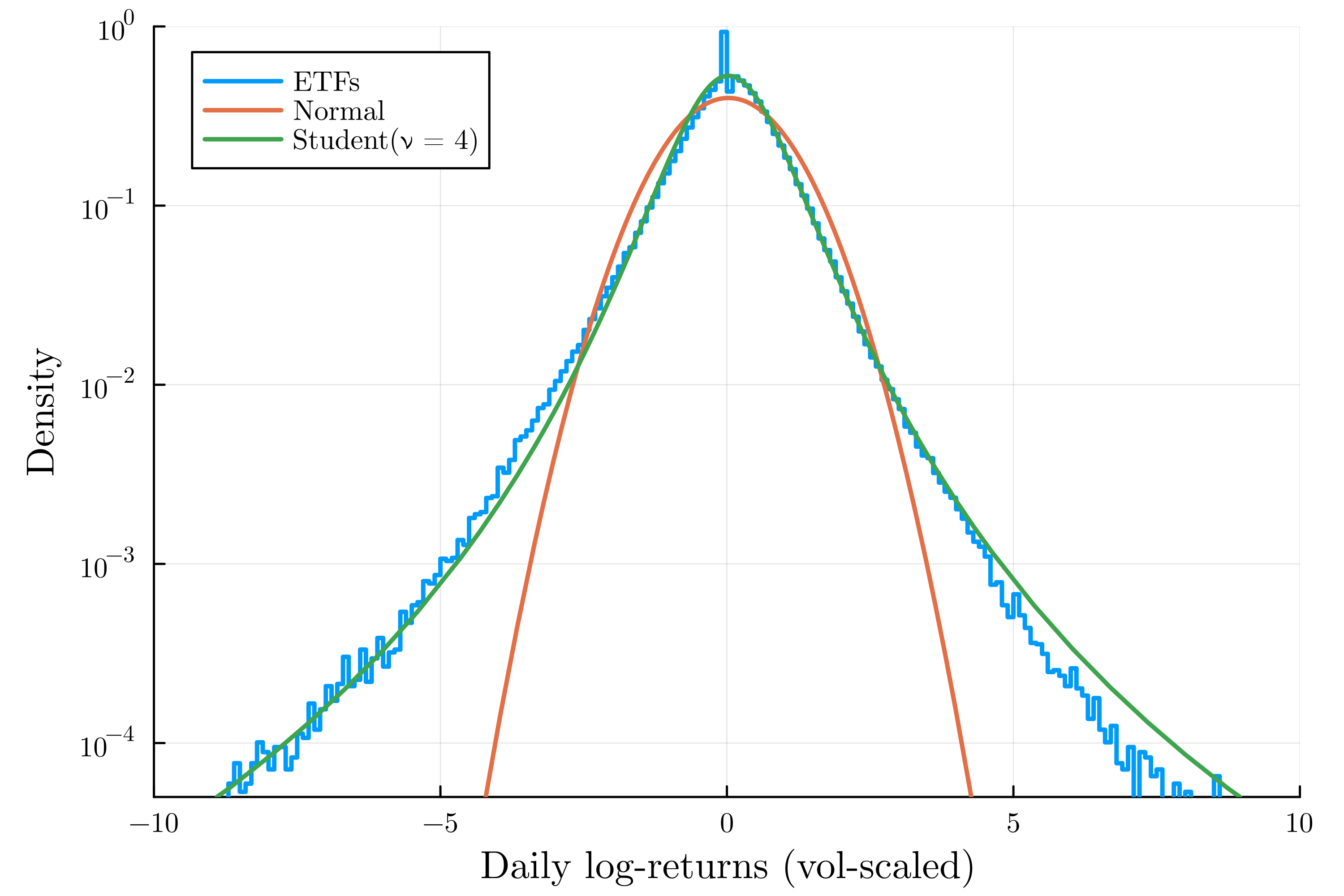}
    \caption{Distribution of daily excess returns (normalized by their average volatility) $\eta_t$ for $1,608$ ETFs, together with Student and Gaussian densities with the same mean and standard deviation.}
    \label{returns-dist}
\end{figure}

The agregate distribution of daily returns of these ETFs (normalized by their long-term volatility $\sigma$) is shown in Fig. \ref{returns-dist}, together with the density of a Student distribution with $\nu =4$ and a Gaussian distribution with the same mean and standard deviation. 
As expected, the Student distribution fits ETF returns well, while the Gaussian clearly does not. 

\subsection{Distribution of Sharpe ratios}

For each of our synthetic distributions (normal and Student with $\nu =3$), we generate $N = 10^5$ samples of $T = 252$ returns and compute their Sharpe ratios $S=\sqrt{252}m/s$. 
The corresponding sampling distribution is plotted in Fig. \ref{returns-dist}, together with the 1-year Sharpe ratios of the $1,608$ ETFs in our sample. 
We also plot for comparison a normal density with mean $\overline{S} = \sqrt{T}\mu/\sigma\simeq 0.13$ and deviation $\Delta S = (1+\overline{S}^2/2)^{1/2}\simeq 1.00$, representing the asymptotic density derived by Lo \cite{loStatistics2002}.

\begin{figure*}
    \includegraphics[width = \textwidth]{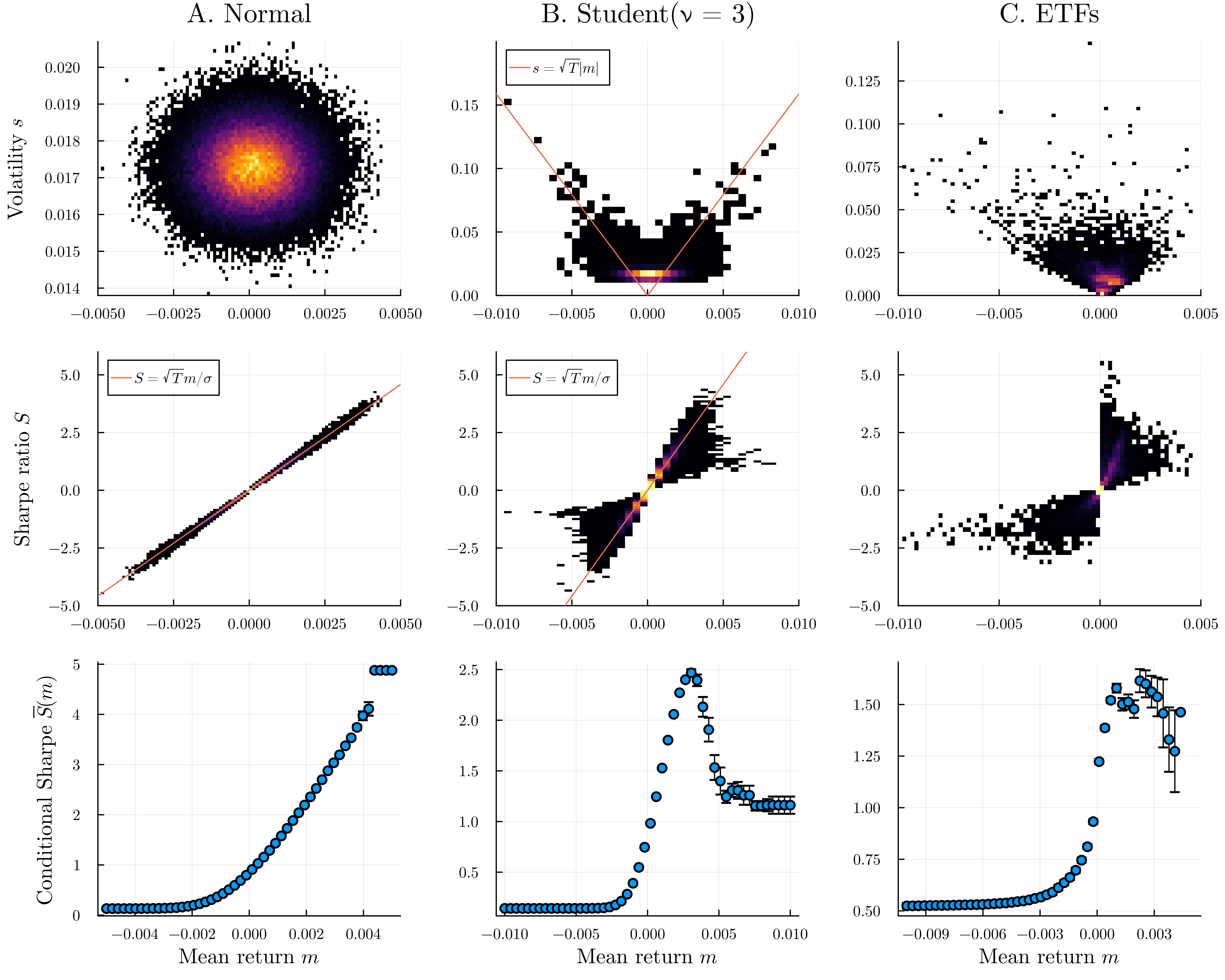}
    \caption{First row: association between mean return $m$ and volatility $s$ when daily returns are drawn from a normal distribution (A), drawn from a Student distribution (B), or computed with real ETF data (C). The color shading indicates the density of points. Second row: the association be the mean return $m$ and Sharpe ratio $S$ is linear in the normal case (A), but non-trivial with heavy-tailed return distributions (B and C). Third row: the non-monotonic nature of the association between performance and Sharpe is revealed by considered the conditional Sharpe ratio $\overline{S}(m)$, defined as the mean Sharpe ratio for funds with mean return at least $m$.}
    \label{main-plot}
\end{figure*}

Fig. \ref{sharpe-dist} shows that asymptotic theory provides a good approximation of the distribution of Sharpe ratios.
More importantly, it shows that sampling noise of the Sharpe ratio is very large, as suggested by its standard error $\Delta S \gg \overline{S}$.

In applications, it is common to grade funds by Sharpe ratios; according to Forbes Advisor, a Sharpe ratio between 1 and 2 is considered good, a ratio between 2 and 3 is very good, and any result higher than 3 is excellent. 
Here we see that the sampling noise of Sharpe ratios covers this range. While the theoretical Sharpe ratio of our synthetic fund is $\mathbb{E}[S] = .13$ (a mediocre value), about $20\%$ of the samples have $S \geq 1$ (``good''), and a non-negligible fraction has $S \geq 2$ (``very good''). 
Clearly, luck plays a big role in deciding whether a Sharpe ratio is ``good" or not. 

\subsection{Joint sampling distributions}

Next we consider the joint sampling distribution of performance (sample mean return $m$), volatility (sample standard deviation $s$), and Sharpe $S$. 

The relevant mathematical fact here is that the sample mean and sample standard deviation of iid samples are independent if and only if these samples are Gaussian \cite{gearyDistribution1936}; in general their joint distribution is non-trivial, with a density given by a multi-dimensional integral \cite{springerJoint1953}. 
The first line of Fig. \ref{main-plot} confirms this fact: while Gaussian funds (A) display no correlation between $m$ and $s$, Student funds (B) and real ETFs (C) exhibit an asymptotic association between $m$ and $s$. 
In the Student case, this association is captured by the asymptotic relation $s \sim \sqrt{T}\vert m\vert$. 
Performance and volatility also exhibit statistical dependence in the ETF case (C), albeit not one that can be easily interpreted.  

Correlations between mean return $m$ and volatility $s$ translate into a non-trivial relationship between mean return $m$ and Sharpe $S$, as illustrated in the second row of Fig. \ref{main-plot}. 
While Gaussian Sharpe ratios are well described by the naive linear relationship $S \sim \sqrt{T}m/\sigma$ (i.e. by fixing the volatility to its expected value), we see in column B and C that this pattern is broken by heavy-tailed returns.
Instead of being maximized by the funds with the largest in-sample performance, the in-sample Sharpe ratio grows and then decreases with $m$. 
This is especially clear if we consider the conditional Sharpe ratio $\overline{S}(m)$, defined as the mean Sharpe ratio of funds with mean return $\geq m$, see third row of Fig. \ref{main-plot}.
Conversely, we find that the funds with largest Sharpe ratio are never the ones with the best realized performance. 

\section{Conclusion}

The Sharpe ratio has some well-known limitations as a measure of investment quality. 
For instance, the Sharpe does not distinguish between upside and downside risk; large tail risks can easily be concealed behind the veil of low volatility, e.g. by writing far out-of-the-money options; most importantly, as we saw, Sharpe ratios consist are largely noise, with standard errors an order of magnitude larger than their values.

None of these caveats relate to the quality of the Sharpe ratio as a descriptor of \emph{realized} returns. 
In this note I have shown that, in constrast with the Gaussian case, a very large Sharpe ratio is not synonymous with great in-sample performance. 
On the contrary, funds with maximal Sharpe ratios tend to have suboptimal performance. 

A few years ago, a Financial Times columnist asked ``If something has a Sharpe Ratio of 8.38, does that mean I should sell my grandmother down the river and buy it?'' \cite{shubberIf2016}. 
Before jumping to hazardous conclusions, it is reasonable to ask: did that thing actually make money? 

\medskip

The Julia code used to generate the present results is freely available at \url{https://github.com/msmerlak/sharpe}. I thank Jean-Philippe Bouchaud for useful critical comments on an earlier version of this manuscript. 

\bibliography{refs.bib}
\end{document}